\documentclass[a4paper,11pt]{article}
\usepackage{latexsym}
\usepackage{epsfig}

\def\bg#1{\mbox{\boldmath$#1$}}

\newcommand{\del}{\partial}
\newcommand{\beq}{\begin{eqnarray}}
\newcommand{\eeq}{\end{eqnarray}}
\newcommand{\be}{\begin{eqnarray*}}
\newcommand{\ee}{\end{eqnarray*}}
\newcommand{\bk}{{\bf k}}

\newcommand{\bx}{{\bf x}}
\newcommand{\om}{\omega}
\newcommand{\Om}{\Omega}

\newcommand{\ra}{\rightarrow}
\newcommand{\e}{\epsilon}

\newcommand{\ket}[1]{\mbox{$\mid\!#1\rangle$}}
\newcommand{\bra}[1]{\mbox{$\langle#1\!\mid$}}

\newcommand{\ex}[1]{\langle\,#1\,\rangle}

\newcommand{\half}{{1\over 2}}

\begin{document}

\centerline{\Large\bf {Scalar Field Fluctuations between Parallel Plates }}
\vskip 5mm
\centerline{Konrad Tywoniuk and Finn Ravndal}
\vskip 5mm
\centerline{\it Department of Physics, University of Oslo, N-0316 Oslo, Norway.}

\begin{abstract}

Quantum fluctuations of a scalar field and its derivatives are calculated when the field is confined between
two parallel plates satisfying Dirichlet or Neumann boundary conditions. After regulation these fluctuations
diverge in general when one approaches one of the plates. The energy density and the pressure between the plates
is only consistent with the total Casimir energy when the canonical energy-momentum tensor is augmented by the Huggins term so
to satisfy the requirement of conformal invariance for a massless, scalar field.

\end{abstract}

\section{Introduction}

Fluctuations in a quantum field can under certain conditions give rise to forces between macroscopic objects.
This was first realized by Casimir\cite{Casimir} who considered an electromagnetic field between two parallel
plates separated by a distance $L$. From the quantum fluctuations of the field described by the Lagrangian
${\cal L}_{EM} = (1/2)({\bf E}^2 - {\bf B}^2)$, he found an attractive force
corresponding to the potential energy
\beq
                    E_0 = - {\pi^2\over 720 L^3}
\eeq
per unit plate area. Assuming a uniform distribution of the energy between the plates, one then has in this region the
energy density
\beq
               {\cal E}_0  = - {\pi^2\over 720 L^4}                                  \label{EM}
\eeq
while outside of the plates one expects it to be zero. 

That it should be so, is not entirely obvious. For the similar Casimir energy inside a spherical shell, explicit
calculations show that the energy density varies with the distance from the shell\cite{OR}, both inside and outside. 
With two parallell plates, the energy density is given by the expectation value of the fluctuations in the electric 
and magnetic fields,
\beq
                 {\cal E} = {1\over 2}\Big[\ex{\bf E^2} + \ex{\bf B^2}\Big]
\eeq
These fluctuations were first calculated by L\"utken and Ravndal\cite{LR} and found to vary between the plates. In fact,
both of them diverge when one of the plates is approached. But this position dependence cancels in the sum, resulting
in a constant energy density.
 
As pointed out by Brown and Maclay\cite{Lowell}, this follows from the more fundamental requirement that the
energy-momentum tensor $T_{\mu\nu}$ of the electromagnetic field is traceless since the photon is massless. With
two parallel plates normal to the unit vector $n^\mu = (0;0,0,1)$ along the $z$-axis, its expectation value must
then have the form
\beq
           \ex{T_{\mu\nu}} = - {\pi^2\over 720 L^4}\big(\eta_{\mu\nu} + 4n_\mu n_\nu \big)
\eeq
with the metric $\eta_{\mu\nu} = \mbox{diag}(1,-1,-1,-1)$. By construction, the energy density is then given by
$\ex{T_{00}}$ and the pressure on the plates is $\ex{T_{zz}} = - \pi^2/240L^4$. 

A detailed calculation of these effects for the electromagnetic field is a bit cumbersome since it must be expanded in
transverse electric and magnetic multipoles. This complication is absent for the massless scalar field $\phi(x)$ which is spinless
and described by the Lagrangian ${\cal L}_0 = \half (\del_\mu\phi)^2 = \half [{\dot\phi}^2 - ({\bg\nabla}\phi)^2]$. In fact, 
it is usually the scalar field which is used when one wants to arrive at the Casimir energy in the simplest way. Not unexpected,
one obtains a total energy which is exactly one half of the electromagnetic result (\ref{EM}) since the photon has two
spin degrees of freedom. The corresponding energy density should then follow from the Hamiltonian density,
\beq
                 {\cal E}^{(0)} = {1\over 2}\Big[\ex{{\dot\phi}^2} + \ex{({\bg\nabla}\phi)^2}\Big]        \label{E_0}
\eeq
These expectation values giving the field fluctuations between the plates will be calculated in the following. As for the 
electromagnetic system, both of the contributions to the energy density vary with the position between the plates. 
However, in this scalar case their sum does not add up to a constant value. 
Instead we find that the sum diverges near the plates. As a consequence, the integrated energy density does not agree with
the known, full energy and is in fact divergent.

The reason for this complication has been known for a long time and is caused by the conformal invariance which is 
present for all free, massless fields and requires a traceless energy-momentum tensor as for the electromagnetic field. In
the above calculation we used the Hamiltonian density which is the $00$-component of the canonical energy-momentum tensor
\beq
          T_{\;\mu\nu}^{(0)} = \del_\mu\phi\del_\nu\phi - {1\over 2}\eta_{\mu\nu}(\del_\lambda\phi)^2       \label{canT}
\eeq
which has the trace $T_{\;\;\mu}^{(0)\mu} = -(\del_\lambda\phi)^2$. The improved energy-momentum tensor derived by 
Callan, Coleman and Jackiw\cite{Callan} has an additional piece
\beq
                 \Delta T_{\mu\nu} = -{1\over 6}(\del_\mu\del_\nu - \eta_{\mu\nu}\del^2)\phi^2             \label{huggins}
\eeq
which had been obtained earlier by Huggins\cite{Huggins}. The full tensor $T_{\mu\nu} =  T_{\mu\nu}^{(0)} +  \Delta T_{\mu\nu}$ 
has zero trace using the equation of motion $\del^2\phi = 0$. We then find a resulting energy density which is
constant and  integrates up to the full energy. The need for this extra term was first pointed out by deWitt who
calculated the full energy-momentum tensor and showed that it gave a
finite Casimir energy\cite{deWitt}. More recently, it has been
calculated by Milton\cite{Milton} using Green's functions 
methods, obtaining the same result for the energy density.

In the next chapter the scalar field between two parrallel plates is quantized in a standard mode expansion for both 
Dirichlet and Neumann boundary conditions. The Casimir energy is calculated making use of combined zeta-function and 
dimensional regularizations to obtain a finite, physical result. With these quantized modes one can then also calculate the
fluctuations in the field and its derivatives. This is done in the following chapter where the same regularization
suffice here to make such local quantities finite. We can then also calculate the energy density and it is
shown that the conformally covariant energy-momentum tensor gives a finite and consistent result both for Dirichlet and 
Neumann boundary conditions. In an appendix we give a short summary of conformal transformations and the derivation of
the correct energy-momentum tensor from the invariant theory in curved spacetime.

\section{Eigenmodes and the Casimir energy}
The massless scalar field $\phi =\phi(\bx,t)$ is defined in a volume closed by two parallel infinite plates.
It is convenient to introduce the coordinate split $\bx=(\bx_T,z)$ where the $\bx_T = (x,y)$ describes the position along 
the surfaces, while the coordinate $z$ describes the position normal to the plates and therefore $z \,\e\, [0,L]$ in the region
between them. 
The dynamics of the classical field is given by the Klein-Gordon field equation  $\del^2\phi = 0$. At the surfaces we will
impose Dirichlet $\phi\vert_{z = 0,L} = 0$ or Neumann $\del_z\phi\vert_{z = 0,L} = 0$ boundary condition. We then have the 
following two sets of orthonormalized eigenmode functions
\beq 
            u_{\bk_T,n} (\bx,t) \;=\; \sqrt{\frac{2}{L}} \bigg(\!\begin{array}{c} \sin(k_{n}z) \\
            \cos(k_{n}z) \end{array}\!\bigg) \, e^{i(\bk_T\cdot\bx_T - \om t)} 
\eeq
where 
\be
       \om = \big(\bk^2_T \,+\, k_{n}^2 \big)^{1/2}
\ee
is the frequency of the mode and $k_{n} = n\pi/L$ with $n = 1,2,3,\ldots$ is the longitudinal wavenumber. The
Neumann mode with $n=0$ is constant in space and doesn't contribute here.

The field can now be expanded in terms of these basic modes
\beq 
         \phi(\bx,t) \;=\; \sum_{n=1}^{\infty} \, \int \frac{d^2k_T}{(2\pi)^2} \, 
         \sqrt{1\over 2\om} \bigg[a_{\bk_T,n}u_{\bk_T,n}(\bx,t) + a^\dagger_{\bk_T,n}u^*_{\bk_T,n}(\bx,t) \bigg] \label{phi}
\eeq 
where the expansion coefficients become creation and annihilation operators after quantization. With the above normalization of the
eigenmodes, they satisfy the canonical commutator
\beq 
      \big[a_{\bk_T,n} , {a^{\dagger}}_{\bk'_T,n'} \big] \;=\; (2\pi)^2 \, \delta(\bk_T - \bk'_T) \, \delta_{n,n'} 
\eeq  
We will here in particular consider the vacuum state $\ket{0}$ defined by the standard condition
$a_{\bk_T,n} \ket{0} = 0$ for all modes specified by the mode quantum numbers $(\bk_T,n)$. 

Summing up the zero-point energies $(1/2)\om$ of each mode, the full vacuum energy per unit plate area is given by
\beq
        E_0 = {1\over 2}\sum_{n=1}^\infty\int\!{d^2k_T\over (2\pi)^2} \Big(\bk^2_T \,+\,(n\pi/L)^2 \Big)^{1/2}
\eeq
In order to make this divergent expression finite, we make use of the basic integral
\beq
      \int\!{d^dk\over (2\pi)^d}{1\over(k^2 + m^2)^N} = {\Gamma(N-d/2)\over (4\pi)^{d/2}\Gamma(N)}(m^2)^{d/2 -N}
\eeq    
from dimensional regularization. This is convergent when the power $N$ is suffiently large compared with the dimension $d$.
When this is not so as in the above integral for the vacuum energy, we define its value by analytical continuation in
the dimension $d$. In our case $d=2$ and $N=-1/2$ which gives
\beq
       E_0 = - {\pi^2\over 12L^3}\sum_{n=1}^\infty n^3                               \label{E_tot}
\eeq
The remaining, divergent sum is done by analytical continuation of the Riemann zeta-function, defined by
\be
              \zeta(s) = \sum_{n=1}^\infty{1\over n^s}
\ee
for $\mbox{Re}\,{s} > 1$. This gives $\sum_{n=1}^\infty n^3 = \zeta(-3) = 1/120$ and the scalar Casimir energy takes 
on the well-known value
\beq
                    E_0 = - {\pi^2\over 1440 L^3}                                      \label{E_full}
\eeq
corresponding to the pressure $p = -\pi^2/480L^4$ normal to the plates. We thus have the same result both for Dirichlet 
and Neumann boundary conditions. In other regularization schemes there sometimes
arise a subtle difference between the two having to do with the zero mode $n=0$ which doesn't contribute in
dimensional regularization\cite{Casimir2}.

\section{Field fluctuations}

Since the field is linear in creation and annihilation operators, the expectation value of the field
is zero in all states. This applies in particular to the vacuum state, $\bra{0}\phi(x)\ket{0} = 0$. But
the fluctuation $\bra{0}\phi^2(x)\ket{0}$ is non-zero. It can be calculated from the field expansion (\ref{phi}) which gives
\be
     \bra{0}\phi^2(x)\ket{0}  &=& \frac{2}{L} \sum_{n,n' = 1}^\infty \, \int\!\int\frac{d^2 k_T}{(2\pi)^2}\frac{d^2k'_T}{(2\pi)^2}\,
     \frac{1}{2\sqrt{\om \om'}}\,\bra{0}a_{\bk_T,n}\,a^{\dag}_{\bk'_T,n'}\ket{0} \\
      &\times& \bigg(\!\begin{array}{c} \sin(k_{n}z)\sin(k_{n'}z)\\ \cos(k_{n}z) \cos(k_{n'}z) \end{array}\!\bigg) \,
     e^{i(\bk_T -\bk'_T)\cdot\bx_T - i(\om-\om')t} \\ \;
      &=&\; \frac{1}{2L} \, \sum_{n=1}^\infty\, \int \frac{d^2 k_T}{(2\pi)^2} \, \frac{1}{\om} \big(1 \,\mp\,
      \cos 2n\theta \big)  
\ee    
where $\theta = \pi z/L$ gives the distance from one plate. Here and in the following upper signs are for Dirichlet 
and lower signs are for Neumann boundary conditions. The momentum integral can now again be made finite with
dimensional regularization as in the previous chapter. One then finds
\beq
      \bra{0}\phi^2(x)\ket{0} = -{1\over 4 L^2} \sum_{n=1}^\infty n (1 \mp \cos 2n\theta)
\eeq 
The first sum here is given by $\sum_{n=1}^\infty n = -1/12$ with zeta-function regularization while the second sum
becomes
\be
      \sum_{n=1}^\infty n\cos 2n\theta = {1\over 2}{\del\over\del\theta}\sum_{n=1}^\infty\sin 2n\theta 
                                       = {1\over 4}\bigg({\del\over\del\theta}\bigg)\cot\theta = - {1\over 4\sin^2\theta}
\ee  
We thus have for the field fluctuation  
\beq
        \bra{0}\phi^2(x)\ket{0} = {1\over 48 L^2}\Big(1 \mp {3\over\sin^2\theta}\Big)
\eeq  
It is seen to diverge near the plates where $\theta\ra 0,\pi$. 

The field fluctuation outside just one plate can now be obatined by removing the other plate to infinity, i.e. letting 
$L\ra \infty$ in the above result. We then find
\beq
       \bra{0}\phi^2(x)\ket{0} = \mp{1\over 16\pi^2 z^2}
\eeq  
It is difficult to imagine any physical measurement of these scalar fluctuations. But the corresponding electromagnetic
field fluctuations will perturb the energy levels of any atom near such a plate and should in principle be observable\cite{LR}.  

\section{Energy-momentum tensors}

From the canonical energy-momentum tensor (\ref{canT}) we have the expression (\ref{E_0}) for the energy density between the 
plates. The first expectation value $\ex{{\dot\phi}^2}$ can be calculated as in the previous chapter. We then obtain
\be
      \bra{0}{\dot\phi}^2\ket{0} = \frac{1}{2L}\,\sum_{n=1}^\infty\,\int \frac{d^2 k_T}{(2\pi)^2}\,\om 
                                   \big(1 \,\mp\,\cos 2n\theta \big)  
\ee    
After dimensional integration over the transverse momenta, which is the same as for the full Casimir energy (\ref{E_tot}), 
we have
\be
      \bra{0}{\dot\phi}^2\ket{0} = - {\pi^2\over 12L^4}\sum_{n=1}^\infty n^3 (1 \mp \cos 2n\theta)
\ee
The remaining sum is similarly found to be
\be
      \sum_{n=1}^\infty n^3\cos 2n\theta =  -{1\over 16}\bigg({\del\over\del\theta}\bigg)^3\!\cot\theta 
                                         \equiv {1\over 8}f(\theta)
\ee  
where the function
\beq
      f(\theta) = {3\over\sin^4\theta} - {2\over\sin^2\theta}
\eeq 
is the same as appears in the derivation of the electromagnetic fluctuations in the same geometry\cite{LR}.
We thus have $\bra{0}{\dot\phi}^2\ket{0} = -(A\mp B)$ with
\beq
         A = {\pi^2\over 12L^4}\sum_{n=1}^\infty n^3 = {\pi^2\over 1440 L^4}
\eeq
and
\beq
         B = {\pi^2\over 12L^4}\sum_{n=1}^\infty n^3\cos 2n\theta =  {\pi^2\over 96L^4}f(\theta)
\eeq
The other needed expectation values can be obtained in the same calculational scheme with the result  
$\bra{0}(\del_z\phi)^2\ket{0}= -3(A \pm B)$ and $\bra{0}({\bg\nabla}_T\phi)^2\ket{0} = 2(A\mp B)$. 
Thus  $\bra{0}(\del_\lambda\phi)^2\ket{0} = \pm 6B$. 

From (\ref{E_0}) we now find ${\cal E}_0 = -(A\pm 2B)$ for the canonical energy density. Due to the presence of the
$B$-term, it is
not constant between the plates and actually diverges when one approaches one of the plates. In the same way we will
find that the expectation value $\ex{T_{zz}}$ does not reproduce the force between the plates. However, let us now
consider the Huggins term (\ref{huggins}). Using the equation of motion $\del^2\phi = 0$, it can be written as
\beq
        \Delta T_{\mu\nu} = -{1\over 3}\Big(\del_\mu\phi\del_\nu\phi 
                          +\phi\del_\mu\del_\nu\phi - \eta_{\mu\nu}(\del_\lambda\phi)^2\Big)           
\eeq
Since $\ex{{\dot\phi}^2} = -\ex{\phi\del_t^2\phi}$, the first two parts cancel for the 00-component which thus becomes
$\ex{\Delta T_{00}} = \pm 2B$ and cancels the position-dependent part of  ${\cal E}_0$. The full, conformally covariant 
energy density is therefore simply ${\cal E} = - A$ as expected from the full energy (\ref{E_full}).

The pressure follows similarly from $zz$-component of the full energy-momentum tensor which can be simplified to
\beq 
      T_{zz} = {2\over 3}(\del_z\phi)^2 - {1\over 3}\phi\,\del_z^2\phi + {1\over 6}(\del_\lambda\phi)^2
\eeq
Notice that $\ex{(\del_z\phi)^2}$ cannot be directly related to $\ex{\phi\,\del_z^2\phi} = 3(A\mp B)$. Putting in the other
expectation values, we find $\ex{ T_{zz}} = - 3A = -\pi^2/480L^4$ which is the presure on the plates.

Needless to say, the conformal Huggins term will also contribute to the energy density and pressure when the system is
in thermal equilibrium at finite temperature. The above vacuum expectation values are then replaced by thermal averages.
These can also be calculated by standard methods to give results on closed form since the fields are free, but this will
not be pursued here.
    
\section{Conclusion}

For the scalar contribution to the Casimir force of two parallel plates the Huggins term is not important since it integrates
out to zero in the total energy. But for the local energy it is essential for consistency. We have seen that the quantum 
fluctuations vary between the plates and actually diverge when one of the plates are approached. It is only with the
conformal Huggins term that the energy density is evenly distributed between the plates and zero outside. 

Scalar fields play an important role in modern cosmology\cite{Pad}. At very early times the inflationary epoch is
driven by the vacuum expectation value of the scalar inflaton field. And at much later time one can model the dark
energy responsible for the acceleration of the cosmic expansion by a scalar quintessence field. In both of these cases
there is the question about what the kinetic energy of the field should be. Quantum effects for a scalar field in a curved 
background will generate a counterterm term $R\phi^2$ which should therefore be included in the Lagrangian. It will generate
a term in the energy-momentum tensor proportional to the Huggins term although there is {\it a priori} no underlying conformal 
invariance in the system. For this reason, such a term should in general be included but with an unknown coefficient.
It will modify the behaviour during the
inflationary epoch\cite{inflaton} and later affect the acceleration of the universe. In fact, recently is has been 
shown that an effective equation of state $p = w\rho$ for the dark energy can result with $w < -1$ which 
represents very strange physics now called 'phantom energy'\cite{Faraoni}. 
A better understanding of these effects will have to wait until we have a deeper theory of fundamental interactions where 
such scalar fields arise naturally. 

\section{Appendix}

Conformal invariance and the energy momentum tensor for scalar fields is discussed in the book by Birrell and Davis\cite{BD}.
We will here give a short summary and a bit more detailed derivation of the Huggins term. 

For a system described by the Lagrangian ${\cal L}$, the action is
\be
               S = \int\!d^4x\sqrt{-g}{\cal L}
\ee 
in curved spacetime with the metric $g_{\mu\nu}$ with $g=\det{g_{\mu\nu}}$. Under an arbitrary variation
$g_{\mu\nu} \ra g_{\mu\nu} + \delta g_{\mu\nu}$ it changes by
\be
        \delta S = \int\!d^4x\Big(\delta\sqrt{-g}\,{\cal L} + \sqrt{-g}\,\delta{\cal L}\Big)
\ee
Since $\delta\sqrt{-g} = -\half\sqrt{-g}g_{\mu\nu}\delta g^{\mu\nu}$ we can write the change in the action as
\beq
        \delta S = {1\over 2} \int\!d^4x\sqrt{-g}T_{\mu\nu}\delta g^{\mu\nu}                     \label{deltaS}
\eeq
where
\beq
        T_{\mu\nu} = 2{\del{\cal L}\over\del g^{\mu\nu}} - g_{\mu\nu}{\cal L}                    \label{erg-mom}
\eeq
is the energy-momentum tensor. For the minimal Lagrangian ${\cal L}^{(0)} = \half g^{\mu\nu}\del_\mu\phi\del_\nu\phi$ this gives
the canonical energy-momentum tensor (\ref{canT}) in flat spacetime.

Under a conformal transformation $g_{\mu\nu}\ra \Om^2g_{\mu\nu}$ where $\Om = \Om(x)$, the metric changes by an amount
$\delta g_{\mu\nu} \propto  g_{\mu\nu}$. If we now demand invariance under such a transformation, we see  from
(\ref{deltaS}) that  $\delta S = 0$ then requires $T_{\;\;\mu}^{\mu} = 0$, i.e. the energy-momentum tensor must be traceless.

Since the scalar field has mass-dimension $d_\phi = 1$ in a 4-dimensional spacetime, it will transform as $\phi \ra \Om^{-1}\phi$. 
In the kinetic energy we will therefore have the
change
\be
            \phi_{,\mu} \ra -\Om^{-2}\Om_{,\mu}\phi +  \Om^{-1}\phi_{,\mu}
\ee
where the comma derivative $\phi_{,\mu}\equiv \del_\mu\phi$. This extra piece generates new terms in the kinetic energy of 
the field which must be cancelled by an additional part in the minimal Lagrangian. We will now show that the extended
Lagrangian
\beq
      {\cal L} = {1\over 2} g^{\mu\nu}\del_\mu\phi\del_\nu\phi   + {1\over 12}R\phi^2                \label{conf_L}
\eeq
where $R= g^{\mu\nu}R_{\mu\nu}$ is the scalar Ricci curvature, is conformally invariant. Since it follows directly
from the definition that $\sqrt{-g}\ra \Om^4 \sqrt{-g}$ under the transformation while for the curvature\cite{BD}
\be
          R \ra \Om^{-2}R - 6\Om^{-3} g^{\mu\nu}\Om_{;\mu\nu}\; ,
\ee  
we see that the full Lagrangian changes into
\be
     \sqrt{-g}{\cal L} \ra   \sqrt{-g} {\cal L} + \sqrt{-g}Q
\ee
The extra term is
\be
     Q = -{1\over 2} g^{\mu\nu}\Big(\big[-\Om^{-2}\Om_{;\mu}\Om_{;\nu} + \Om^{-1}\Om_{;\mu\nu}\big]\phi^2 
       + 2\phi\phi_{;\nu}\Om^{-1}\Om_{;\mu}\Big)
\ee 
after having changed all comma derivatives of scalar quantities into covariant derivatives. Now we can simplify this expression 
to the total derivative $Q = -\half(\Om^{-1}\Om^{;\mu}\phi^2)_{;\mu}$ which gives a vanishing contribution in the integral 
for the action. The Lagrangian (\ref{conf_L}) is therefore invariant under conformal transformations.

Under the variation $g_{\mu\nu} \ra g_{\mu\nu} + \delta g_{\mu\nu}$ we see that the last term in the invariant Lagrangian
(\ref{conf_L}) will generate a new contribution to the energy-momentum tensor from the additional term
\be
        \delta S_R =  {1\over 12}\int\!d^4x\Big(\delta\sqrt{-g}\,R + \sqrt{-g}\,\delta g^{\mu\nu}R_{\mu\nu} 
                   +  \sqrt{-g}\, g^{\mu\nu}\delta R_{\mu\nu}\Big)\phi^2
\ee
in the variation of the action. Here we will need the variation of the Ricci tensor which can be expressed in terms of the 
Christoffel symbols as
\beq
        R_{\mu\nu} = \Gamma^{\alpha}_{\mu\nu,\alpha} -  \Gamma^{\alpha}_{\mu\alpha,\nu} 
                   + \Gamma^{\alpha}_{\mu\nu} \Gamma^{\beta}_{\alpha\beta} 
                   - \Gamma^{\beta}_{\mu\alpha}\Gamma^{\alpha}_{\nu\beta} 
\eeq
The calculation is most easily performed in Gaussian normal
coordinates, where the Christoffel symbols are zero, but their
derivatives are non-zero. We then have
\be
    g^{\mu\nu}\delta R_{\mu\nu} =  g^{\mu\nu}\big(\delta\Gamma^{\alpha}_{\mu\nu,\alpha}  
                                -  \delta\Gamma^{\alpha}_{\mu\alpha,\nu}\big) 
                                =  \big(g^{\mu\nu}\delta\Gamma^{\alpha}_{\mu\nu} 
                                -  g^{\mu\alpha}\delta\Gamma^{\nu}_{\mu\nu} \big)_{,\alpha}  
\ee 
since the derivatives of the metric is zero in these coordinates. The expression within the parenthesis is a vector $w^\alpha$ since
$\delta\Gamma^{\alpha}_{\mu\nu} = g^{\alpha\beta}(\delta g_{\beta\mu,\nu} - \half\delta g_{\mu\nu,\beta})$ is a tensor.  
Together with $\delta\Gamma^\nu_{\mu\nu} = \half g^{\nu\beta}\delta g_{\nu\beta,\mu}$ we then have for this vector
\beq
        w^\alpha = -  g^{\mu\nu}g^{\alpha\beta}[\delta g_{\mu\nu,\beta} - \delta g_{\beta\mu,\nu}]
\eeq
Collecting terms, we can write the new contribution to the variation as
\be
         \delta S_R = {1\over 12}\int\!d^4x\sqrt{-g}\Big(E_{\mu\nu}\delta g^{\mu\nu} 
                    - g^{\mu\nu}g^{\alpha\beta}[\delta g_{\mu\nu,\alpha\beta} - \delta g_{\beta\mu,\alpha\nu}]\Big)\phi^2
\ee
where $E_{\mu\nu} = R_{\mu\nu} - \half g_{\mu\nu}R$ is the Einstein tensor. In the last term we perform two partial integrations
resulting in
\be
     \delta S_R = {1\over 12}\int\!d^4x\sqrt{-g}\Big(E_{\mu\nu}\phi^2 
                - \Big[(\phi^2)_{,\mu\nu} -  g_{\mu\nu}(\phi^2)^{,\alpha}_{\;\; ,\alpha}\Big]\Big)\delta g^{\mu\nu}  
\ee 
Comparing with the definition (\ref{deltaS}) of the energy-momentum tensor, we see that we now can identify the general 
Huggins term. In a general coordinate system where partial derivatives $\del_\mu$ are replaced with covariant derivatives 
$\nabla_\mu$, it is
\beq
       \Delta T_{\mu\nu} = {1\over 6}E_{\mu\nu}\phi^2 -  {1\over 6}\Big(\nabla_\mu\nabla_\nu - g_{\mu\nu}\Box\Big)\phi^2
\eeq
where $\Box = \nabla^\mu\nabla_\mu$ is the d'Alembertian operator. In Minkowski spacetime it simplifies to (\ref{huggins})
in the text.

The full energy-momentum tensor
\be
       T_{\mu\nu} =  \del_\mu\phi\del_\nu\phi - {1\over 2}g_{\mu\nu}(\del_\lambda\phi)^2 + \Delta T_{\mu\nu}
\ee
is now traceless,
\be
       T^\mu_{\;\;\mu} &=& (\del_\mu\phi)^2 - 2(\del_\mu\phi)^2 + {1\over 6}(R - 2R)\phi^2 - {1\over 6}(\Box - 4\Box)\phi^2\\
                       &=& - (\del_\mu\phi)^2 - {1\over 6}R\phi^2 + {1\over 2}\Box \phi^2 \\
                       &=& - (\del_\mu\phi)^2 - {1\over 6}R\phi^2 + (\del_\mu\phi)^2 + \phi\Box\phi = 0
\ee
since $\Box\phi = (1/6)R\phi$ is the equation of motion for the conformally invariant Lagrangian (\ref{conf_L}). An additional
massterm $\propto \phi^2$ breaks the invariance, while a $\propto\phi^4$ coupling preserves it and thus will give rise to a 
traceless energy-momentum tensor for the interacting field.


\begin{thebibliography}{99}

\bibitem{Casimir} H.B.G. Casimir, {\it Proc. K. Ned. Akad. Wet.} {\bf 51}, 793 (1948).

\bibitem{OR} K. Olaussen and F. Ravndal, {\it Nucl. Phys.} {\bf B192} 237, (1981).

\bibitem{LR} C.A. L\"utken and F. Ravndal, {\it Phys. Rev.} {\bf A31}, 2082 (1985).

\bibitem{Lowell} L.S. Brown and G.J. Maclay, {\it Phys. Rev.} {\bf 184}, 1272 (1969).

\bibitem{Callan} C.G. Callan, S. Coleman and R. Jackiw, {\it Ann. Phys. N.Y.} {\bf 59}, 42 (1972).

\bibitem{Huggins} E. Huggins, {\it Ph.D. thesis}, Caltech, 1962
  (unpublished).

\bibitem{deWitt} B. deWitt, {\it Phys. Rep.} {\bf 19C}, 295 (1975).

\bibitem{Milton} K.A. Milton, {\it Phys. Rev.} {\bf D68}, 065020 (2003).

\bibitem{Casimir2} F. Ravndal, {\it hep-ph/0009208}.

\bibitem{BD} N.D. Birrell and P.C.W. Davies, {\it Quantum fields in curved space}, Cambridge University Press, 1982.
               
\bibitem{Pad} T. Padmanabhan, {\it Phys. Rep.} {\bf 380}, 235 (2003).

\bibitem{inflaton} F. Lucchin, S. Matarrese and M.D. Pollock, {\it Phys. Lett.} {\bf B167}, 163 (1986).

\bibitem{Faraoni} V. Faraoni, {\it Phys. Rev.} {\bf D68}, 063508 (2003).

\end{thebibliography}
\end{document}